\documentclass[aps,prb,preprint,superscriptaddress,showpacs,showkeys]{revtex4}

\usepackage{graphicx}

\begin{document}

\title{On the dc Magnetization, Spontaneous Vortex State and Specific Heat in the superconducting
state of the weakly ferromagnetic superconductor
RuSr$_{2}$GdCu$_{2}$O$_{8}$ }

\author{Thomas P. Papageorgiou}
\email[Corresponding author:]{T.Papageorgiou@fz-rossendorf.de}
\affiliation{Hochfeld-Magnetlabor Dresden, Forschungszentrum
Rossendorf, D-01314 Dresden, Germany}

\author{Ennio Casini}
\affiliation{Physikalisches Institut, Universit\"at Bayreuth,
D-95440 Bayreuth, Germany}
\author{Hans F. Braun}
\affiliation{Physikalisches Institut, Universit\"at Bayreuth,
D-95440 Bayreuth, Germany}

\author{Thomas Herrmannsd\"orfer}
\affiliation{Hochfeld-Magnetlabor Dresden, Forschungszentrum
Rossendorf, D-01314 Dresden, Germany}
\author{Andrea D. Bianchi}
\affiliation{Hochfeld-Magnetlabor Dresden, Forschungszentrum
Rossendorf, D-01314 Dresden, Germany}
\author{Joachim Wosnitza}
\affiliation{Hochfeld-Magnetlabor Dresden, Forschungszentrum
Rossendorf, D-01314 Dresden, Germany}

\date{\today}

\begin{abstract}

Magnetic-field changes $<$ 0.2 Oe over the scan length in
magnetometers that necessitate sample movement are enough to create
artifacts in the dc magnetization measurements of the weakly
ferromagnetic superconductor RuSr$_{2}$GdCu$_{2}$O$_{8}$ (Ru1212)
below the superconducting transition temperature $T_{c} \approx$ 30
K. The observed features depend on the specific magnetic-field
profile in the sample chamber and this explains the variety of
reported behaviors for this compound below $T_{c}$. An experimental
procedure that combines improvement of the magnetic-field
homogeneity with very small scan lengths and leads to artifact-free
measurements similar to those on a stationary sample has been
developed. This procedure was used to measure the mass magnetization
of Ru1212 as a function of the applied magnetic field H (-20 Oe
$\le$ H $\le$ 20 Oe) at $T < T_{c}$ and discuss, in conjunction with
resistance and ac susceptibility measurements, the possibility of a
spontaneous vortex state (SVS) for this compound. Although the
existence of a SVS can not be excluded, an alternative
interpretation of the results based on the granular nature of the
investigated sample is also possible. Specific-heat measurements of
Sr$_{2}$GdRuO$_{6}$ (Sr2116), the precursor for the preparation of
Ru1212 and thus a possible impurity phase, show that it is unlikely
that Sr2116 is responsible for the specific-heat features observed
for Ru1212 at $T_{c}$.
\end{abstract}



\maketitle

\section{\label{Intro}Introduction}
The high-temperature superconducting cuprate
RuSr$_{2}$GdCu$_{2}$O$_{8}$ (Ru1212), discovered \cite{Bauernfeind}
in 1995, has been the subject of intense investigations in the last
10 years, especially after it was shown that in this compound
superconductivity develops in an already magnetically ordered state.
\cite{Bauernfeind2} The published results were often contradicting
and made it difficult to obtain a clear picture about the magnetic
and superconducting properties of Ru1212. The magnetic properties of
Ru1212 in its normal state will be sketched only briefly here.
Concerning this issue, we would like to notice that at present, it
seems to be widely accepted that the Ru (Gd) moments in
superconducting Ru1212 order antiferromagnetically,\cite{Lynn} at
$T^{Ru}_{M} = T_{M} \approx$ 130 K ($T^{Gd}_{M} \approx$ 2 K), in a
canted arrangement that gives rise to a net magnetic moment.
\cite{Jorgensen} Thus, Ru1212 is magnetically characterised as weak
ferromagnet.

The definition of widely accepted superconducting properties for
Ru1212 was proven much more difficult. It is known that the
preparation conditions significantly affect the superconducting
properties (such as the superconducting transition temperature
$T_{c}$) of Ru1212 and elaborate investigations \cite{Bauernfeind2,
Shaou} had to be untertaken in order to define the preparation route
that leads to good quality samples. Nevertheless, even for good
quality samples, the reported behaviors in dc magnetization (e.g.
Refs.~\onlinecite{Bernhard, Klamut, Artini, Cimberle}) and
specific-heat \cite{Tallon,Chen} measurements below $T_{c}$ differ
significantly and complicate the elucidation of the superconducting
state in Ru1212.

In a series of investigations,\cite{Papageorgiou, Papageorgiou2,
Papageorgiou3, Papageorgiou4, Papageorgiou5} we have tried to
determine the origin of the reported different behaviors in the dc
magnetization measurements of Ru1212 below $T_{c}$. We attributed
them to artifacts arising from the movement of the sample in an
inhomogeneous magnetic field during the measurement in the
Superconducting Quantum Interference Device (SQUID) magnetometer.
Measuring the volume susceptibility of a stationary Ru1212 sample in
a home-made SQUID system,\cite{Papageorgiou4} we were led to the
conclusion, that measurements in magnetometers that do not
necessitate sample movement would lead to the revelation of a
universal behavior in the superconducting state of Ru1212 for
samples of comparable quality. New experimental and software
options, not available for our previous investigations,
\cite{Papageorgiou, Papageorgiou2, Papageorgiou3, Papageorgiou4,
Papageorgiou5} allow us in the present paper to extend this
discussion and provide experimental evidence that strongly supports
this point of view. Additionally, we briefly describe how the
magnetic-field homogeneity in a commercial SQUID system can be
improved and combined with small scan lengths to give artifact-free
measurements similar to those on a stationary sample.

Conditions of high magnetic-field homogeneity (field change
$\approx$ 0.01 Oe over the scan length) have been achieved for the
measurement of the mass magnetization of Ru1212 at $T < T_{c}$ as a
function of the applied magnetic field. These measurements were
combined with resistance and ac susceptibility measurements to
investigate the possibility of a spontaneous vortex state (SVS)
formation in the superconducting state of Ru1212. The fact that
magnetic order is already developed at $T > T_{c}$ could lead to the
creation of vortices, when the sample enters the superconducting
state, even in the absence of external magnetic field, if the
internal magnetic field is greater than $H_{c1}$. Recently, dc
magnetization measurements were combined with magnetoresistance
measurements to construct the phase diagram of Ru1212. The existence
of a SVS has been suggested.\cite{Yang} We show that an alternative
explanation based on the granular nature of the investigated samples
is also possible.

Finally, we address the problem of impurity phase formation in
Ru1212 and its effect on the specific-heat measurements of this
compound. We concentrate our attention on Sr$_{2}$GdRuO$_{6}$
(Sr2116), a compound with interesting magnetic properties below 40 K
\cite{Papageorgiou} used as a precursor for the preparation of
Ru1212.\cite{Bauernfeind2, Shaou} The fact that Sr2116 orders
magnetically in the temperature range where Ru1212 becomes
superconducting \cite{Papageorgiou} has been used as an argument
(e.g. Ref.~\onlinecite{Chu}) to exclude the possibility of bulk
superconductivity in Ru1212; it is proposed that the reported
\cite{Tallon,Chen} anomaly in the specific heat of Ru1212 could be
attributed to the magnetic transition of Sr2116 impurities rather
than the superconductivity of Ru1212. Here, we present specific-heat
measurements of a Sr2116 sample and investigate this possibility.

\section{\label{Exp}Experimental}

Details about the preparation and characterization in terms of X-ray
powder diffraction of the granular (polycrystalline) Ru1212 and
Sr2116 samples can be found in a previous work.\cite{Papageorgiou}

The presented dc magnetization and ac susceptibility measurements
were performed in a Quantum Design magnetic properties measurement
system (MPMS). For the resistance and specific-heat measurements the
physical property measurement system (PPMS) of the same manufacturer
was used. Both systems are part of the equipment of the new Dresden
High Magnetic Field Laboratory (pulsed 100 T / 10 ms project).
\cite{Herrmannsdoerfer,site}

\section{\label{Res}Results and discussion}
\subsection{\label{dc}dc magnetization}

We have described in detail how a standard SQUID magnetometer works
and how artifacts arise in the dc magnetization measurements of
Ru1212 below $T_{c}$.
\cite{Papageorgiou2,Papageorgiou3,Papageorgiou5} Here, we will
expand this discussion and we will provide experimental evidence
that (i) the magnitude of magnetic-field inhomogeneities in the
superconducting magnet of the magnetometer that leads to artifacts
below $T_{c}$ is smaller than 0.2 Oe, (ii) the reported different
behaviors in the dc magnetization measurements of Ru1212 below
$T_{c}$ could be the result of different magnetic-field profiles in
the magnet.

A key point for reliable dc magnetization measurements in the
superconducting state of Ru1212 is to avoid moving the sample in an
inhomogeneous magnetic field.
\cite{Papageorgiou2,Papageorgiou3,Papageorgiou5} We were able to
avoid sample movement in a home-made magnetometer.
\cite{Papageorgiou4} These measurements \cite{Papageorgiou4} will
serve as a reference for comparison in the present study. All dc
magnetization measurements discussed here were performed on the same
Ru1212 sample investigated with the home-made magnetometer.

\begin{figure}
\includegraphics[width=85mm]{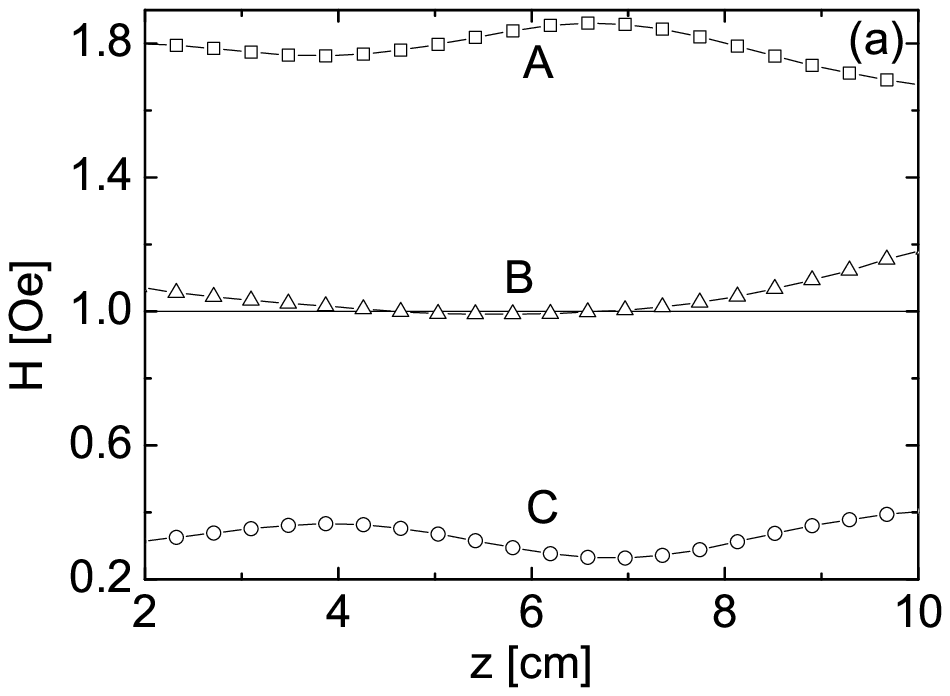}
\includegraphics[width=85mm]{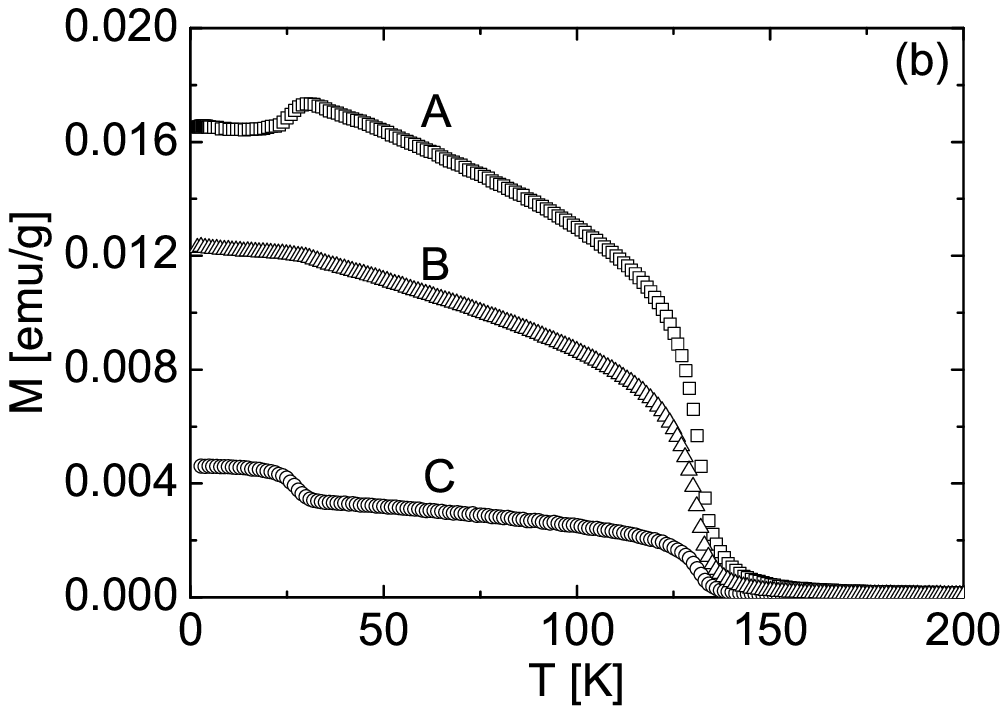}
\caption{\label{fig1} (a) Measured magnetic-field profiles along the
axis of the superconducting magnet in the MPMS. The center of the
magnet is located at z = 6 cm. (b) Mass magnetization measurements
of Ru1212 in the magnetic-field profiles A, B and C, respectively.
All measurements were collected with increasing temperature after
the sample was field-cooled.}
\end{figure}

The magnetic-field profiles A and C shown in Fig.~\ref{fig1}(a),
measured along the axis of the superconducting magnet using the
profile-field operation of the MPMS, were produced after the
following procedures and field changes were utilized; profile A: 70
kOe $\rightarrow$ 0 Oe $\rightarrow$ degauss shield $\rightarrow$
magnet reset $\rightarrow$ 1 Oe, profile C: -70 kOe $\rightarrow$ 0
Oe $\rightarrow$ degauss shield $\rightarrow$ magnet reset
$\rightarrow$ 1 Oe. The degauss shield option of the MPMS involves
the demagnetization of the permalloy shield around the
superconducting magnet. The magnet reset option leads to the
reduction, via a controlled quench, of the residual field in the
superconducting magnet. Nevertheless, the average final values of
the magnetic field differ considerably from the nominal (set) value
of 1 Oe. Profile A is closer to 1.8 Oe and profile C closer to 0.3
Oe. The maximum field change for profile A over the 8 cm distance
around the center of the magnet shown in Fig.~\ref{fig1}(a) is
$\approx$ 0.17 Oe, whereas for profile C it is $\approx$ 0.13 Oe. It
is interesting to note that the two field profiles are almost
symmetric with respect to the final set value of the magnetic field
of 1 Oe.

The mass magnetization of the Ru1212 sample measured with a scan
length of 8 cm is shown in Fig.~\ref{fig1}(b). Curves A and C of
Fig.~\ref{fig1}(b) correspond to the magnetic-field profiles A and C
of Fig.~\ref{fig1}(a), respectively. In the normal state of the
sample, both measurements show the magnetic ordering of the Ru
moments at $T_{M} \approx$ 130 K. The measured ordered moments below
$T_{M}$ are different since the average applied magnetic field for
the two measurements was different, as shown in Fig.~\ref{fig1}(a).
Just below $T_{c} \approx$ 30 K, curve A shows a decrease of the
magnetization followed by a plateau at lower temperatures, whereas
curve C shows a ``symmetric'' effect with an increase of the
magnetization close to $T_{c}$. Both curves are different from those
measured with the sample kept stationary. \cite{Papageorgiou4} This
fact illustrates clearly that field changes $<$ 0.2 Oe over the scan
length are sufficient to create artifacts below $T_{c}$. This is in
very good agreement with an earlier estimate \cite{Papageorgiou3}
based on the Bean-model.\cite{Bean1,Bean2}

\begin{figure}
\includegraphics[width=85mm]{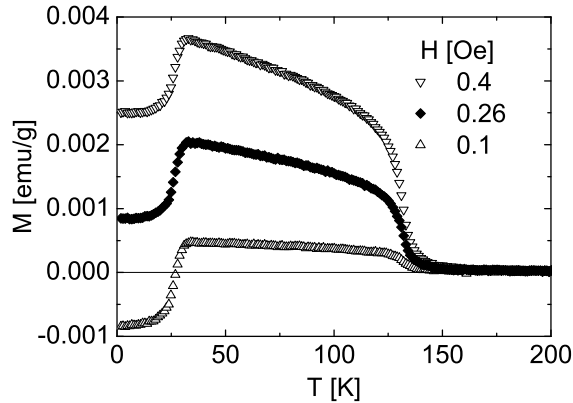}
\caption{\label{fig2} Field-cooled magnetization of the Ru1212
sample measured in magnetic-field profiles similar to profile A of
Fig.~\ref{fig1}(a). The magnetic field values given are those
measured at the center of the magnet.}
\end{figure}

From the measurements presented it is also obvious that different
magnetic-field profiles ``create'' different features in the dc
magnetization measurements of Ru1212 below $T_{c}$. The
reproducibility of a certain feature cannot be considered as
evidence for reliable measurements. Rather, it may be just the
effect of a reproducible field-profile. The shape of the
magnetic-field profile is determined by large field changes similar
to those which led to the profiles A and C in Fig.~\ref{fig1}(a) (70
kOe to 0 Oe and -70 kOe to 0 Oe, respectively). Smaller field
changes, of the order of a few Oe, simply shift the profile without
changing it. This was verified for the field profile A of
Fig.~\ref{fig1}(a) and is shown in Fig.~\ref{fig2}. The small
magnetic-field changes result in a series of measurements with
similar features, which still do not represent the real behavior of
the sample below $T_{c}$. From the above discussions it becomes
obvious that the reported magnetization peculiarities of Ru1212
below $T_{c}$ could be the result of different measuring
magnetic-field profiles. We note that curve A of Fig.~\ref{fig1}(b)
and the curves of Fig.~\ref{fig2} are very similar to those reported
in Ref.~\onlinecite{Bernhard}, while curve C of Fig.~\ref{fig1}(b)
resembles results presented in Ref.~\onlinecite{Klamut}.

\begin{figure}
\includegraphics[width=85mm]{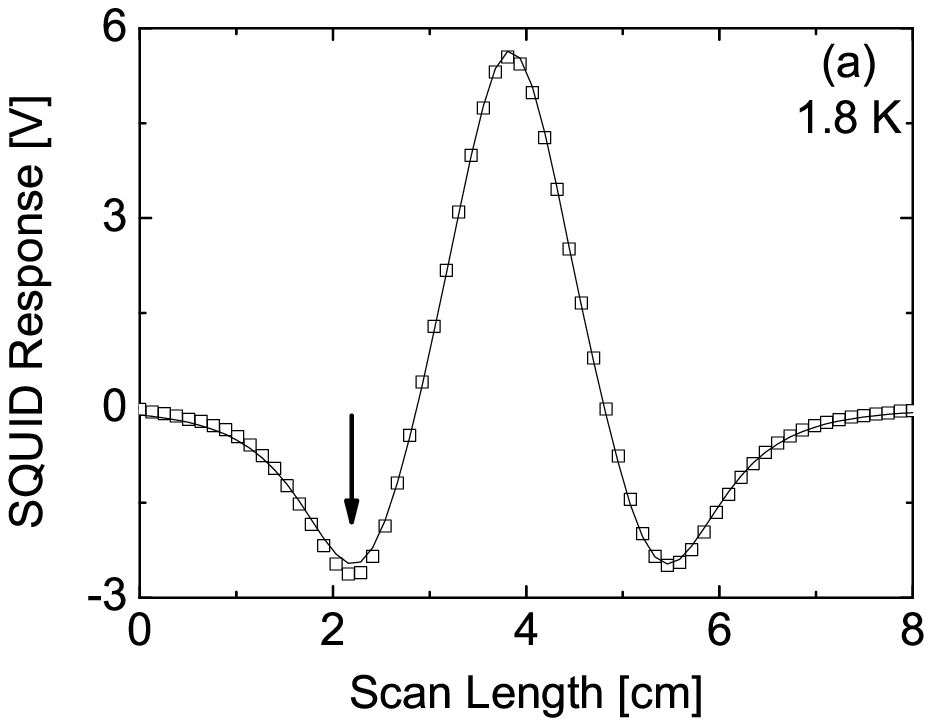}
\includegraphics[width=85mm]{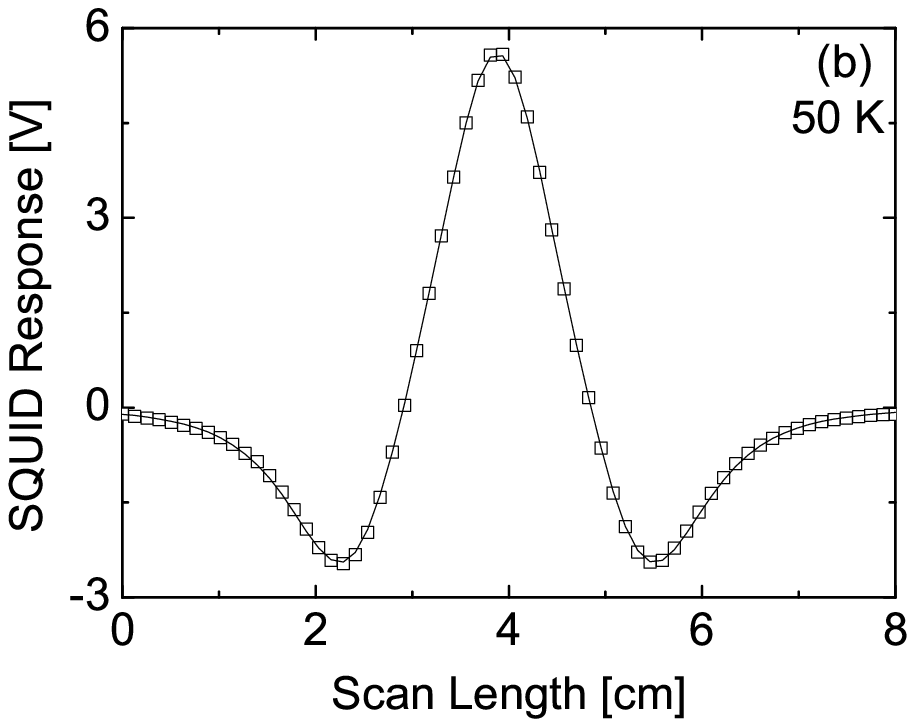}
\caption{\label{fig3} (a) SQUID response to the movement of the
Ru1212 sample over the magnetic-field profile A of
Fig.~\ref{fig1}(a) at 1.8 K (open squares) and the corresponding fit
(solid line) generated from the magnetometer's software for the
calculation of the magnetic moment. (b) The same at 50 K.}
\end{figure}

It should be pointed out, that the problems related with the SQUID
magnetometry of Ru1212 below $T_{c}$ are difficult to notice. In
Fig.~\ref{fig3}, we show the measured SQUID signals (open squares)
and the corresponding fits (solid lines) from which the magnetic
moments at 1.8 and 50 K for curve A of Fig.~\ref{fig1}(b) were
calculated. In the normal state of the sample (50 K), where all
research groups report qualitatively similar results, there is no
distortion of the measured signal and there is a very good agreement
with the fit generated by the magnetometer's software. At first
sight, this appears to be the case also in the superconducting state
(1.8 K), but a closer look of Fig.~\ref{fig3}(a) reveals that the
fit is insufficient close to the minimum on the left hand side of
the curve [arrow in Fig.~\ref{fig3}(a)]. We note that the fits are
done assuming that the magnetic moment of the sample remains
constant over the scan length, which obviously is not true when
magnetic-field inhomogeneities are present. This is an important
point related with the generation of artifacts below $T_{c}$, as
discussed in
Refs.~\onlinecite{Papageorgiou2,Papageorgiou3,Papageorgiou5}.

From the above discussions it becomes obvious, that the improvement
of the magnetic-field homogeneity in the superconducting magnet of
the magnetometer is essential for reliable measurements of Ru1212
samples below $T_{c}$. The SQUID system available for the present
work combines several options for the improvement of the field
homogeneity at low magnetic fields: The degauss shield and magnet
reset options were briefly discussed earlier. An additional option
is the ultra low field option (ULFO), which allows (i) nulling of
the magnetic field at a user-defined position in the sample space
and (ii) measuring of the magnetic field as a function of position
along the axis of the 70 kOe superconducting magnet.

The degauss shield, magnet reset and ULFO (in this order) were
successively used to achieve a field close to 1 Oe over the scan
length. For this, before starting the nulling operation of the ULFO,
an offset of -1 Oe was applied to the magnetic field sensor and the
center of the magnet was chosen as the position, where the field
should be nulled. After this procedure, the magnetic-field profile
over a distance of 8 cm along the axis of the magnet and close to
its center was measured. This measurement, curve B in
Fig.~\ref{fig1}(a), showed a maximum field change $\approx$ 0.01 Oe
in the area defined by 5 cm $<$ z $<$ 7 cm. For smaller distances of
about 0.5 cm around the center of the magnet the magnetic field was
practically constant. Thus, measuring with a very small scan length
close to the center of the magnet should lead to artifact-free
results.

The reciprocating sample option (RSO) of the MPMS was utilized for
the measurements with small scan lengths. This option combines a
servo motor with a digital signal processor and allows more rapid
and accurate measurements. The servo motor, unlike the standard
stepper motor, does not stop sample movement for each data reading.
The sample is oscillating with a user-defined frequency.
Furthermore, sample centering with respect to the pick-up coil
system (second order gradiometer located at the center of the
superconducting magnet) for small scan lengths is easier and
automated with the RSO.

We have used the RSO with a scan length of only 0.5 cm close to the
center of the field profile B of Fig.~\ref{fig1}(a) to measure the
magnetization of our Ru1212 sample. Sample centering took place
above $T_{c}$. The measurement is shown in Fig.~\ref{fig1}(b) (curve
B). Curve B shows, as expected, a rapid increase of the
magnetization close to 130 K related with the (canted)
antiferromagnetic ordering of the Ru moments at this temperature, as
discussed in section~\ref{Intro}. Below T$_{c} \approx$ 30 K neither
a decrease, consistent with field expulsion from the sample, nor an
increase, due to the paramagnetic contribution from the Gd moments,
of the magnetization is observed. Instead, a shallower slope of the
magnetization is observed, indicating competition between
superconductivity and magnetic order. Exactly the same behavior was
observed in Ref.~\onlinecite{Papageorgiou4}. Thus, the experimental
procedure developed above, which combines improvement of the
magnetic-field homogeneity in the superconducting magnet of the
SQUID system with small measuring scan lengths in the area with the
highest field-homogeneity, allows artifact-free measurements below
$T_{c}$ equivalent to those for a stationary sample.
\cite{Papageorgiou4}

It should not be considered that small scan lengths alone can lead
to reliable measurements in the presence of magnetic-field
gradients. The RSO with a scan length of 0.5 cm around the center of
the magnet was also used, after large magnetic-field changes,
without utilizing any of the options that allow improvement of the
field homogeneity. In this case, the measured field profiles showed
a field change $>$ 0.1 Oe close to the center of the magnet and
artifacts in the measurements appeared again.

\subsection{Investigation of possible spontaneous vortex state
formation}

The possible formation of a SVS in Ru1212 has always been an
intriguing question, in view of the fact, that in this compound weak
ferromagnetism coexists with superconductivity. Most recently, the
formation of a SVS has been proposed for Ru1212, after the phase
diagram of this compound has been constructed using magnetization
and magnetoresistance measurements.\cite{Yang} The resistance
measurements in zero applied magnetic field showed a $T_{c,onset}$ =
56 K and $T_{c}$($R$=0) = $T_{c}$ = 39 K. Magnetization measurements
as a function of low applied magnetic field (denoted in the
following as $M(H)$ measurements) below $T_{c}$ were used to
determine the Meissner region in the phase diagram. The average
$H_{c1}$ for each temperature was defined from the peaks of the
initial diamagnetic $M(H)$ curves. For zero applied magnetic field
the Meissner region was proposed to end at T = 30 K $<$ T$_{c}$ = 39
K. This led the authors to the suggestion that a SVS exists between
30 K and $T_{c,onset}$ (vortex glass of lattice phase at 30 K $< T <
T_{c}$ and vortex liquid phase at $T_{c} < T < T_{c,onset}$).

There are, generally, two problems related with $M(H)$ measurements
of Ru1212 below $T_{c}$. The first has to do with the difference
between set and real values of the magnetic field. The second
problem is related with the generation of artifacts in the
measurements, when field inhomogeneities are present over the scan
length. Both problems were illustrated in section~\ref{dc}.

It was shown that improvement of the magnetic-field homogeneity
combined with small scan lengths lead to reliable field-cooled
magnetization measurements of the Ru1212 sample as a function of
temperature. For $M(H)$ measurements at a constant temperature below
$T_{c}$ though, the problem is that the improvement of the field
homogeneity cannot be attempted for every magnetic field. The
application of the ULFO, for example, requires removement of the
sample from the sample chamber. The ULFO can be applied only for the
initial field nulling, before the sample is zero field cooled to the
desired temperature and the measuring magnetic fields are applied.
Nevertheless, we have tried to estimate the magnetic-field profiles
in our sample chamber in the following way: the ULFO was used to
null the magnetic field at the center of the superconducting magnet.
Then the magnetic field was set to 5 Oe $\rightarrow$ 10 Oe
$\rightarrow$ 15 Oe $\rightarrow$ 20 Oe $\rightarrow$ 0 Oe
$\rightarrow$ -5 Oe $\rightarrow$ -10 Oe $\rightarrow$ -15 Oe
$\rightarrow$ -20 Oe $\rightarrow$ 0 Oe. In every step of this
cycle, the magnetic field profile on the magnet axis was measured.
It was found that the average difference H$_{real}$-H$_{set}$ was
+0.67 Oe. The average field change over a distance of 0.5 cm around
the center of the magnet was $\approx$ 0.01 Oe. Under these
conditions, reliable magnetization measurements of Ru1212 below
$T_{c}$ can be performed, as described previously.

\begin{figure}
\includegraphics[width=85mm]{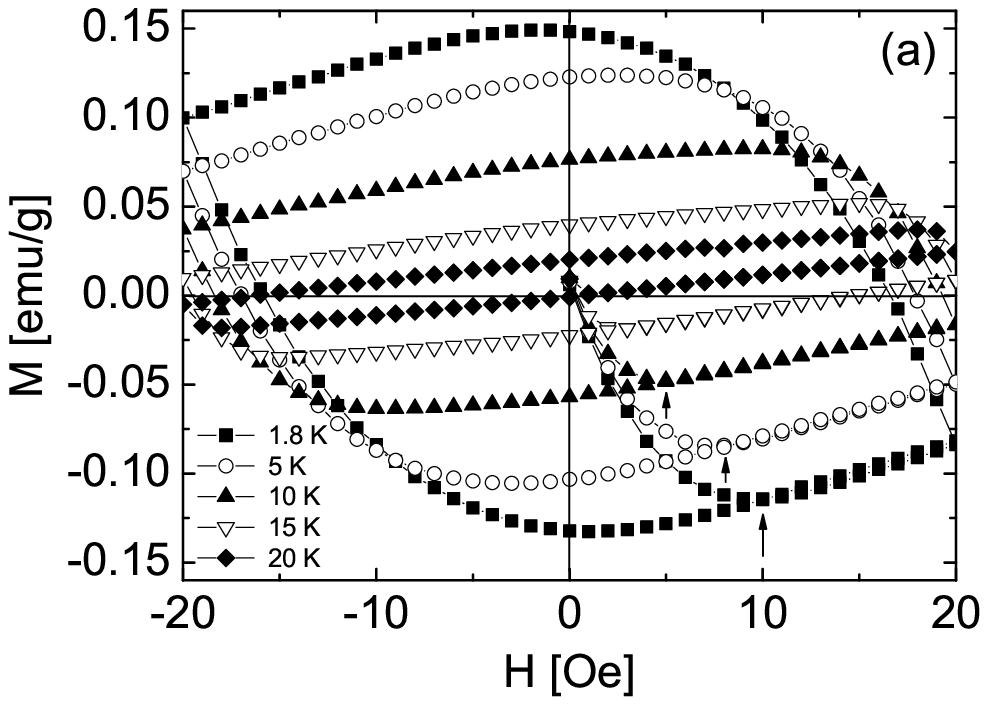}
\includegraphics[width=85mm]{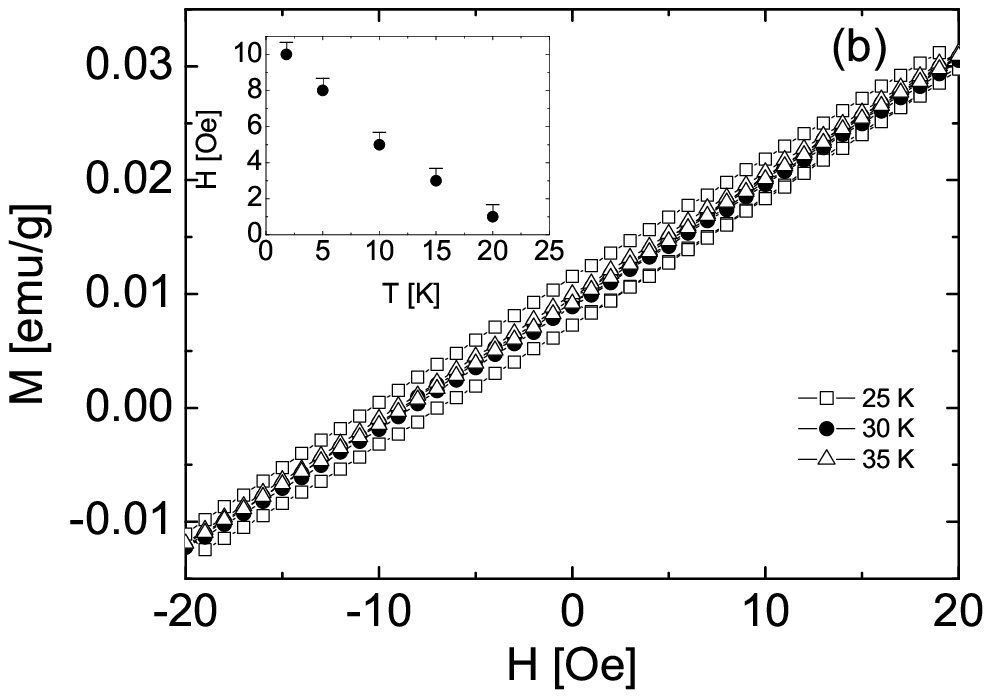}
\caption{\label{fig4} Magnetization of the Ru1212 sample at low
applied magnetic fields for T = 1.8 K, 5 K, 10 K, 15 K, 20 K (a) and
T = 25 K, 30 K, 35 K (b). Inset: $H(T)$ curve obtained using the
magnetic field at which the peak of the initial diamagnetic $M(H)$
curve was observed for each measuring temperature. Some of these
peaks are marked with arrows in Fig.~\ref{fig4}(a). The error bars
correspond to the 0.67 Oe average difference between set and real
value of the magnetic field (see text).}
\end{figure}

Thus, we have obtained $M(H)$ (-20 Oe $\le H \le$ 20 Oe)
measurements of the Ru1212 sample measuring as described previously
with a very small scan length of only 0.5 cm around the center of
the magnet at $T \leq T_{c}$ . The $M(H)$ curves are shown in
Fig.~\ref{fig4}. Contrary to what is reported in
Ref.~\onlinecite{Yang}, the shape of the curves is consistent with
what is expected for the Bean critical-state model \cite{Chen2}
(plus a paramagnetic contribution) and no fluctuations are observed
for the curves measured close to $T_{c}$. Furthermore, as shown in
the inset of Fig.~\ref{fig4}(b), the $H(T)$ curve obtained using the
peaks of the initial diamagnetic $M(H)$ curves, some of which are
marked with arrows in Fig.~\ref{fig4}(a), does not follow a
parabola, as in Ref.~\onlinecite{Yang}, where it was used to define
the Meissner region of the phase diagram of Ru1212.

\begin{figure}
\includegraphics[width=85mm]{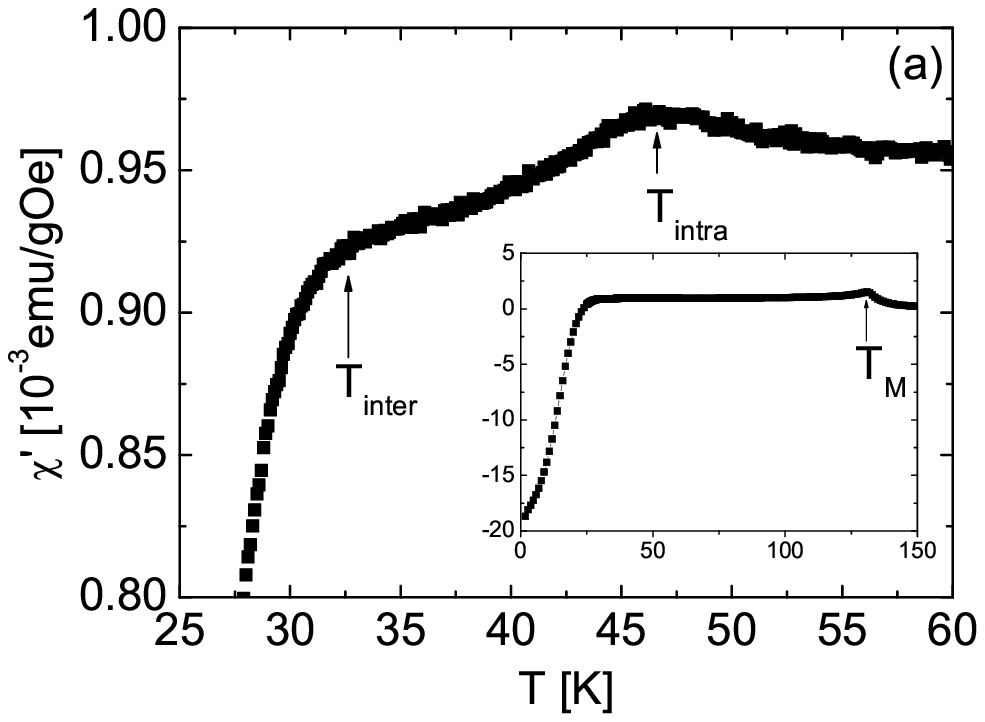}
\includegraphics[width=85mm]{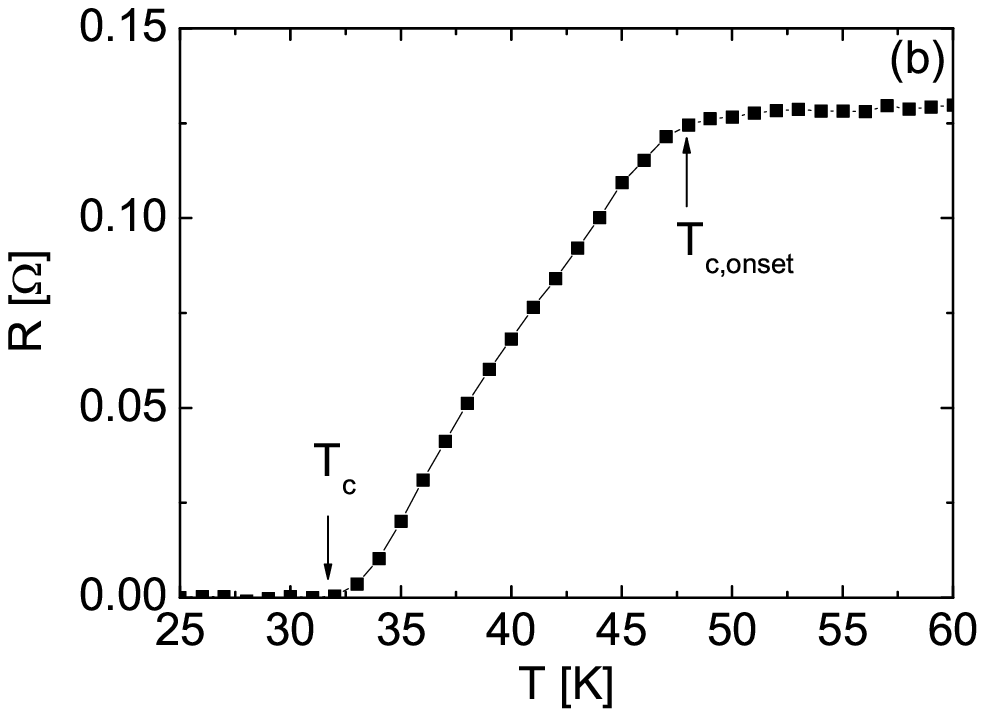}
\caption{\label{fig5} (a) Low-temperature details of the real part
of the ac susceptibility of Ru1212. The excitation-field amplitude
was 3.9 Oe and the frequency 22.2 Hz. No dc magnetic field was
applied. Inset: The whole measurement between 1.8 and 150 K. (b) Low
temperature resistance measurement of the same sample in zero
applied magnetic field.}
\end{figure}

The observed differences lead us to suggest an alternative
interpretation of the available data based on the granular nature of
the investigated samples. In Fig.~\ref{fig5}(a), we show an ac
susceptibility measurement of our Ru1212 sample. The main part of
the figure shows low-temperature details of the real part, $\chi$',
of the ac susceptibility. The inset of  Fig.~\ref{fig5}(a) shows the
whole measurement between 1.8 and 150 K, where apart from the
superconducting transition, the magnetic transition at $T_{M}
\approx$ 130 K is also visible. At low temperatures, there are two
distinct points, where $\chi$' is decreasing with different slopes.
The first is located at $T_{intra} \approx$ 46 K and the second at
$T_{inter} \approx$ 32 K. Similar features, in
Ref.~\onlinecite{Lorenz}, were attributed to the intragrain
($T_{intra}$) and intergrain ($T_{inter}$) transitions of the
granular sample . According to a model developed by J. R. Clem
\cite{Clem}, granular high-temperature superconductors can be viewed
as an array of Josephson-coupled superconducting grains. $T_{intra}$
is the temperature, where the grains become superconducting and
start to shield the magnetic field. This shielding is expressed with
the shallow decrease of $\chi$' below $T_{intra}$. The magnetic
field remains in the intergrain area (area between the grains). At
some lower temperature $T_{inter}$, the coupling between the grains
is established and the whole sample shields the magnetic field. This
explains the rapid decrease of $\chi$' below $T_{inter}$.

In Fig.~\ref{fig5}(b), the resistance of our Ru1212 sample measured
in zero applied magnetic field is shown. Our sample shows
T$_{c,onset} \approx$ 48 K and T$_{c}$($R$=0) = T$_{c}$ = 32 K,
somewhat lower compared to the sample studied in
Ref.\onlinecite{Yang}. The width of the transition though ($\approx$
16 K) is almost the same. It is interesting to note that
T$_{c,onset}$ fits nicely with T$_{intra}$ and $T_{c} \approx
T_{inter}$. From this comparison we conclude that the resistive part
of the superconducting transition between $T_{c,onset}$ and $T_{c}$
could be the result of resistive contacts between the grains and not
necessarily of a spontaneous vortex liquid phase.

In Ref.~\onlinecite{Yang} though, there is a temperature interval
between 30 K and $T_{c}$ = 39 K, outside the Meissner region of the
suggested phase diagram of Ru1212, where the resistance is zero in
zero applied magnetic field. The authors suggest a spontaneous
vortex glass or lattice state for this temperature range. In this
case, one has to consider where the vortices are pinned. A granular
superconductor is characterised by two characteristic lower critical
fields $H_{c1}$: $H_{c1J}$ and $H_{c1g}$. At $H_{c1J}$ the magnetic
field starts to penetrate between the grains and at $H_{c1g}$ into
the grains, with $H_{c1J} < H_{c1g}$. If the vortices are created
and pinned in the intergrain area, then the SVS would be related to
the granular nature of the sample. It would be highly desirable to
test the proposed SVS in bulk single crystalline Ru1212 samples.

\subsection{Specific Heat}

Contradicting reports concerning the superconducting properties of
Ru1212 are not related only with dc magnetization measurements but
also with specific-heat measurements for this compound below
$T_{c}$. A typical example are the papers by Tallon et
al.\cite{Tallon} and Chen et al.\cite{Chen} The most striking
difference between these reports is that, whereas in the former an
increase of $T_{c}$ is reported with the application of an external
magnetic field, raising the possibility for triplet pairing in
Ru1212, in the latter, a more ``conventional'' picture is given,
with a decrease of $T_{c}$, when magnetic field is applied. Another
difference is that Tallon et al.\cite{Tallon} could not directly
observe specific-heat features attributed to the superconductivity
of Ru1212. In this study \cite{Tallon} a differential heat-capacity
measurement was utilised with a 3 $\%$ Zn-substituded
non-superconducting Ru1212 sample as reference. Chen et
al.\cite{Chen} could directly observe the reported specific heat
peaks below $T_{c}$.

In a previous work, \cite{Papageorgiou} we have presented evidence
of magnetic ordering of both the Ru and Gd moments in the compound
Sr$_{2}$GdRuO$_{6}$ (Sr2116) at about 30 and 3 K, respectively.
Sr2116 was used as a precursor for the preparation of the Ru1212
samples on which the specific-heat measurements were performed.
\cite{Tallon,Chen} Chu et al. argue that this compound (Sr2116)
could be responsible, as an impurity phase, for the specific-heat
features observed for Ru1212 below $T_{c}$.\cite{Chu} In the
following, we will discuss this possibility.

\begin{figure}
\includegraphics[width=85mm]{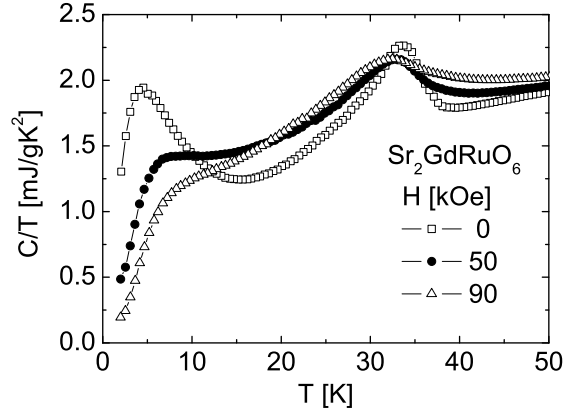}
\caption{\label{fig6} Specific heat of Sr$_{2}$GdRuO$_{6}$ measured
in different magnetic fields.}
\end{figure}

Fig.~\ref{fig6} shows the specific heat, C/T, of a Sr2116 sample
measured with the PPMS in different magnetic fields up to 90 kOe.
The curve measured in 0 Oe shows two peaks at 34 K and 4.5 K,
respectively related to the magnetic-ordering effects reported
previously. \cite{Papageorgiou} We will concentrate our attention to
the peak observed at 34 K, since it is close to $T_{c}$ of Ru1212.
The size of this peak, C/T(34 K)-C/T(40 K), is $\approx$ 0.47
mJ/gK$^{2}$ at $H$ = 0 and is obviously decreasing with the
application of a magnetic field. The position of the peak is
slightly shifting from 34 K at 0 Oe to 32 K at 90 kOe.

The behavior described above cannot account for the findings of
Tallon et al.\cite{Tallon} Nevertheless, an increase of $T_{c}$ by
4.5 K, when the magnetic field is increased from 0 to 40-50 kOe
should be obvious, for example, in resistance measurements as well.
In Ref.~\onlinecite{Yang} though, $T_{c}$($R$=0) is reduced from 39
K, in zero external magnetic field, to 16 K in a field of 50 kOe. It
is also reported \cite{Yang}, that $T_{c,onset}$, which is related
to the thermodynamic transition temperature, taking into account the
granular nature of the samples, is also reduced by 3 K, when a
magnetic field of 70 kOe is applied. On the other hand, Lorenz et
al. \cite{Lorenz} report for the sister compound
RuSr$_{2}$EuCu$_{2}$O$_{8}$ an increase of the temperature, where
the resistance drop towards zero begins, when a magnetic field is
applied. The possibility of triplet pairing for Ru1212 calls for
further experimental investigations.

The specific-heat peaks observed for Sr2116 close to 34 K show a
magnetic-field dependence similar to that reported by Chen et
al.\cite{Chen} for Ru1212. Actually, in Ref.~\cite{Chen} it is
stated that Sr2116 impurities were detected in the investigated
samples. Even so, it is difficult to attribute the features observed
for Ru1212 to Sr2116 impurities. The size of the peak measured in
zero field for the Ru1212 sample was \cite{Chen} $~\approx$ 0.08
mJ/gK$^{2}$. Assuming a Ru1212 sample mass of 50 mg and taking into
account the size of the Sr2116 peak in zero field given above, about
16$\%$ Sr2116 impurities would be required to explain the Ru1212
peak of Ref.~\onlinecite{Chu}. We expect that such an amount of
impurity would create more than one weak peak on the X-ray pattern
of Ru1212, as it is stated in Ref.~\cite{Chen} It should be noted
though that evidence has been presented,\cite{Papageorgiou} that the
stoichiometry or doping state (e.g. with Cu) of Sr2116 could affect
its magnetic properties. This would have definetely an effect also
in the measured specific heat. Since the stoichiometry or doping
state of possible Sr2116-like impurities in Ru1212 is difficult to
be defined, the above comparison should be considered as a rough and
not a definite one. The argument that Chen et al.\cite{Chen} use
against the possibility that the specific-heat features observed for
Ru1212 are due to Sr2116 impurities is that a non-superconsucting
Ru1212 sample obtained by annealing the superconducting one in
flowing nitrogen at 500 $^{\circ}$C for 48 h did not show any
specific-heat anomaly between 20 and 50 K. A heat treatment in
nitrogen could indeed increase the amount of Sr2116 impurities in
the sample. It is actually known, that Ru1212 decomposes to Sr2116
and Cu$_{2}$O, when heated in nitrogen at temperatures around 1000
$^{\circ}$C. \cite{Bauernfeind3}

\section{Concluding remarks}

We have shown that for magnetometers that necessitate sample
movement, magnetic-field inhomogeneities of less than 0.2 Oe in the
superconducting magnet are enough to create artifacts in the dc
magnetization measurements of Ru1212 below T$_{c}$. In this case,
the observed features depend on the specific field profile in the
magnet and this explains the variety of reported unusual behaviors
for Ru1212 below $T_{c}$. Measurements in magnetometers, where the
sample remains stationary, are the most reliable ones and will, with
very high probability, reveal a universal behavior for Ru1212
samples of similar quality. For magnetometers that necessitate
sample movement, artifact-free measurements are also possible, but
special care with respect to the magnetic-field homogeneity has to
be taken (section~\ref{dc}).

The formation of a spontaneous vortex state for Ru1212 can not be
excluded. Nevertheless, the granularity of the investigated samples
has to be carefully taken into account. A particular vortex state
with vortices pinned in the intergrain area is much more likely.
Single crystals would be required to unambiguously demonstrate the
formation or non-formation of a spontaneous vortex state in bulk
Ru1212.

Specific-heat measurements of Sr2116, the precursor for the
preparation of Ru1212, showed anomalies in the temperature range,
where Ru1212 becomes superconducting. A comparison with the reported
specific-heat measurements of Ru1212 led us to the conclusion that
it is rather unlikely that Sr2116, as an impurity phase, is
responsible for the specific-heat anomalies observed in the
superconducting state of Ru1212. The elucidation of the pairing
mechanism in Ru1212 though, requires further investigations.

\bibliography{Paper1}

\end{document}